\documentclass[12pt]{article}

\usepackage{amsmath,amssymb,amsthm}
\usepackage{amsmath,amssymb}

\newcommand{\n}{\noindent}

\begin{document}
\def\a{\alpha}\def\b{\beta}\def\c{\chi}\def\d{\delta}\def\e{\epsilon}
\def\f{\phi}\def\g{\gamma}\def\h{\theta}\def\i{\iota}\def\j{\vartheta}
\def\k{\kappa}\def\l{\lambda}\def\m{\mu}\def\n{\nu}\def\o{\omega}\def
\p{\pi}\def\q{\psi}\def\r{\rho}\def\s{\sigma}\def\t{\tau}\def\u{\upsilon}
\def\vu{\varphi}\def\w{\varpi}\def\y{\eta}\def\x{\xi}\def\z{\zeta}

\def\D{\Delta}\def\F{\Phi}\def\G{\Gamma}\def\H{\Theta}\def\L{\Lambda}
\def\O{\Omega}\def\P{\Pi}\def\Q{\Psi}\def\S{\Sigma}\def\U{\Upsilon}\def\X{\Xi}

\def\lie{{\cal L}}\def\de{\partial}\def\na{\nabla}\def\per{\times}
\def\inf{\infty}\def\id{\equiv}\def\mo{{-1}}\def\ha{{1\over 2}}
\def\qu{{1\over 4}}\def\pro{\propto}\def\app{\approx}
\def\circa{\sim}\def\we{\wedge}\def\di{{\rm d}}\def\Di{{\rm D}}

\def\bg{background }\def\gs{ground state }\def\bhs{black holes }
\def\des{de Sitter }
\def\ades{(anti)-de Sitter }\def\QM{quantum mechanics }
\def\cor{commutation relations }\def\up{uncertainty principle }

\def\section#1{\bigskip\noindent{\bf#1}\smallskip}
\def\subsect#1{\bigskip\noindent{\it#1}\smallskip}
\def\nota{\footnote{$^\dagger$}}
\font\small = cmr8

\def\PL#1{Phys.\ Lett.\ {\bf#1}}\def\CMP#1{Commun.\ Math.\ Phys.\ {\bf#1}}
\def\PRL#1{Phys.\ Rev.\ Lett.\ {\bf#1}}\def\AP#1#2{Ann.\ Phys.\ (#1) {\bf#2}}
\def\PR#1{Phys.\ Rev.\ {\bf#1}}\def\CQG#1{Class.\ Quantum Grav.\ {\bf#1}}
\def\NP#1{Nucl.\ Phys.\ {\bf#1}}\def\GRG#1{Gen.\ Relativ.\ Grav.\ {\bf#1}}
\def\JMP#1{J.\ Math.\ Phys.\ {\bf#1}}\def\PTP#1{Prog.\ Theor.\ Phys.\ {\bf#1}}
\def\PRS#1{Proc.\ R. Soc.\ Lond.\ {\bf#1}}\def\NC#1{Nuovo Cimento {\bf#1}}
\def\JoP#1{J.\ Phys.\ {\bf#1}} \def\IJMP#1{Int.\ J. Mod.\ Phys.\ {\bf #1}}
\def\MPL#1{Mod.\ Phys.\ Lett.\ {\bf #1}} \def\EL#1{Europhys.\ Lett.\ {\bf #1}}
\def\AIHP#1{Ann.\ Inst.\ H. Poincar\'e {\bf#1}}\def\PRep#1{Phys.\ Rep.\ {\bf#1}}
\def\AoP#1{Ann.\ Phys.\ {\bf#1}}\def\AoM#1{Ann.\ Math.\ {\bf#1}}
\def\JHEP#1{JHEP\ {\bf#1}}\def\RMP#1{Rev.\ Mod.\ Phys.\ {\bf#1}}
\def\grq#1{{\tt gr-qc/#1}}\def\hep#1{{\tt hep-th/#1}}\def\arx#1{{\tt arXiv:#1}}

\def\ref#1{\medskip\everypar={\hangindent 2\parindent}#1}
\def\beginref{\begingroup}
\bigskip
%\centerline{ References}
\nobreak{\noindent}
\def\endref{\par\endgroup}

\def\bx{\bar x}\def\bp{\bar p}\def\br{\bar r}
\def\mcr{modified commutation relations }\def\tran{transformations }
\def \schr{Schr\"odinger }\def\ads{anti-de Sitter }\def\rep{representation }
\def\cor{commutation relations }\def\up{uncertainty principle }

%%%%%%%%%%%%%%%%%%%%%%%%%%%%%%%%%%%%%%%%%%%%%%%%%%%%%%%%%%%%%%%%%%%%%%%%%%%%%%%%%%%%%%%%
%\magnification=1200
\baselineskip=18pt
%{\nopagenumbers}
%\line{\hfil
%March 2010
%}
%\vskip30pt
\vskip10pt
\centerline{\bf Quantum mechanics in de Sitter space}
\vskip40pt
\centerline{
{\bf Subir Ghosh}$^a$
and
{\bf Salvatore Mignemi}$^b$}
\vskip10pt
\centerline{$^b$ Physics and Applied Mathematics Unit, Indian Statistical Institute}
\centerline{203 B.T. Road, Kolkata 700108, India}

\centerline{$^b$ Dipartimento di Matematica, Universit\`a di Cagliari}
\centerline{viale Merello 92, 09123 Cagliari, Italy}
\centerline{and INFN, Sezione di Cagliari}
\vskip60pt
\centerline{\bf Abstract}

\vskip10pt
{\noindent
We consider some possible phenomenological implications of the extended uncertainty
principle, which is believed to hold for quantum mechanics in de Sitter spacetime.
The relative size of the corrections to the standard results is however of the order
of the ratio between the length scale of the quantum mechanical system and the \des
radius, and therefore exceedingly small. Nevertheless, the existence of effects due
to the large scale curvature of spacetime in atomic experiments has a theoretical
relevance.
}
\vskip60pt\
P.A.C.S. Numbers: 04.60.-m; 03.65.Ta
\vfil{\eject}

\section{1. Introduction}

The possibility that quantum gravity induces the deformation of the \cor of \QM has been
widely discussed in recent years [1]. The corrections should be proportional to the square
of the Planck length $l_P\sim10^{-35}$ m.
For example, one may assume [2]\footnote{We use units such that $\hbar=c=1$.}
$$\D x_i\D p_j\ge{\d_{ij}\over2}\left(1+l_P^2\D p_i^2\right).\eqno(1)$$
A relation of this kind has been called generalized uncertainty principle (GUP).

The implications of this hypothesis on quantum mechanical systems have been considered
in several papers [3-5]. Corrections of relative size $l_P^2/a^2\sim10^{-48}$, where
$a=1/me^2$ is the Bohr radius, are for example expected to arise in the spectrum of the
hydrogen atom [4,5].
Although these effects are too small to be experimentally detectable, observations can
fix limits on the value of the deviations from the Heisenberg formula.

On the other hand, similar effects can also derive from more classical settings.
For example, it has been argued that in a \des background the Heisenberg
uncertainty principle should be modified by introducing corrections
proportional to the cosmological constant $\L=3\l^2$ [6],
$$\D x_i\D p_j\ge{\d_{ij}\over2}\left(1-\l^2\D x_i^2\right).\eqno(2)$$
This modification of the Heisenberg relation was named extended uncertainty principle
(EUP).
It has been motivated either by analogy with the GUP, or by gedanken experiments in
which the expansion of
the universe during a measurement is taken into account [7]. More recently, it has
been shown that it can also be derived from the definition of \QM on a \des background,
with a suitably chosen parametrization [8].

In this letter, we discuss the implications of the EUP on quantum mechanical
systems, in analogy with the investigation made in refs.\ [3-5] for the GUP. In
particular, we define through a nonlinear transformation new variables, that obey
canonical commutation relations, and calculate perturbatively their effect on the
spectrum of the harmonic oscillator and of the hydrogen atom.

\section{2. Perturbations of the spectrum}

The spatial part of the deformed Heisenberg algebra leading to the extended uncertainty
principle, studied in [8], is given by
$$[x_i,x_j]=0,\qquad[p_i,p_j]=-\l^2J_{ij},\qquad[x_i,p_j]=\d_{ij}-\l^2x_ix_j,\eqno(3)$$
where $J_{ij}=x_ip_j-x_jp_i$. The \up (2) follows from (3) if $<x_i>\ =0$. For spherical
symmetric systems this is true for all states, provided the origin of the coordinates is
put in the center of symmetry, which is always possible because of the homogeneity of
\des spacetime.

A \rep of the \cor (3) can be obtained from operators $\bx_i$ and $\bp_i$
satisfying canonical commutation relations, through the nonlinear \tran [9]
$$x_i={\bx_i\over\sqrt{1+\l^2\bx^2}},\qquad p_i=\sqrt{1+\l^2\bx^2}\ \bp_i.\eqno(4)$$
In particular, in a position \rep $\bx_i$ acts as a multiplication operator, while
$\bar p_i=-i{\de\over\de\bar x_i}$.
The \schr equation for the variables $x_i$, $p_i$,
$$\left[{p^2\over2m}+V(x)\right]\q=E\,\q,\eqno(5)$$
can then be obtained by substituting (4) into (5). Since the exact form of the \tran (4)
is not easy to handle, we shall consider an expansion at first order in the small
parameter $\l^2$.
In the following, we shall consider central potentials for which $V=V(r)$, with
$r=\sqrt{x_i^2}$. We shall therefore expand
$$r={\br\over\sqrt{1+\l^2\br^2}}\sim(1-{\l^2\over2}\,\br^2)\br,\qquad
p^2=\left(1+\l^2\br^2\right)\bp^2,\eqno(6)$$
and the \schr equation becomes at first order in $\l^2$,
$$\left[{\bp^2\over2m}+V(\br)+{\l^2\over2}\left(\br^2\,{\bp^2\over m}-\br^3\,{dV(\br)\over
d\bar r}\right)\right]\q\id\left[\bar H+{\l^2\over2}\D H\right]\q=E\,\q,\eqno(7)$$
where $\bar H$ is the original Hamiltonian, but written in terms of the barred operators.
We can consider $\D H$ as a small perturbation to the Hamiltonian $\bar H$.
Notice that the first term in $\D H$ may give rise to ordering problems. When
necessary we shall adopt the symmetric ordering $\ha(\br^2\bp^2+\bp^2\br^2)$.

The spectrum of the Hamiltonian $H$ can be obtained through perturbation theory as
$$E_k=\bar E_k+{\l^2\over2}\D E_k,\eqno(8)$$
where $k$ denotes the energy levels of the unperturbed Hamiltonian $\bar H$, and
$\D E_k$ are the eigenvalues of the matrix
$$<k\,|\D H|\,k'>\ =\ \left<k\left|\br^2\,{\bp^2\over m}-\br^3\,{dV\over d\bar r}\right|k'
\right>,\eqno(9)$$
calculated on degenerate states of the given energy level.
\bigbreak
\subsect{2.1. Harmonic oscillator}

The simplest example is given by a three-dimensional harmonic oscillator. Its Hamiltonian
is
$$H={p^2\over2m}+{k\over2}\,r^2.\eqno(10)$$
Its energy eigenvalues are parametrized by the quantum numbers
$n$ and $l$, while the quantum number $m$ does not enter the calculations:
$$\bar E_{n,l}=\sqrt{k\over m}\left(2n+l+{3\over2}\right).\eqno(11)$$
The perturbation is at first order,
$$\D H={\br^2\bp^2\over m}-k\br^4,\eqno(12)$$
and the calculation of the energy shift is simplified by the use of the identity [4]
$${\bp^2\over 2m}=\bar H-{k\over2}\,\br^2,\eqno(13)$$
which gives
$$<n,l\,|\D H|\,n,l>\ =2\bar E_{n,l}<n,l\,|\,\br^2|\,n,l>-2k<n,l\,|\,\br^4|\,n,l>.\eqno(14)$$
The matrices $<n,l\,|\,\br^2|\,n,l>$ and $<n,l\,|\,\br^4|\,n,l>$ are diagonal, and an
explicit calculation gives [4]
$$<n,l\,|\,\br^2|\,n,l>\ =\ {\bar E_{n,l}\over2k},$$
$$<n,l\,|\,\br^4|\,n,l>\ =\ {1\over km}\left(6n^2+9n+6nl+l^2+4l+{15\over4}\right).\eqno(15)$$
The shift in the energy levels due to the extended \up is therefore
$$\D E_{n,l}=-{1\over m}\left(8n^2+12n+8nl+l^2+5l+{21\over4}\right).\eqno(16)$$
For $l=0$, the relative magnitude of the corrections is then
$${\l^2\over2}\,{\D E_{n,0}\over E_{n,0}}\sim-\left(n+{3\over2}\right){\l^2\over\sqrt{km}}\,.
\eqno(17)$$

\subsect{2.2. Hydrogen atom}

The Hamiltonian of the hydrogen atom is
$$H={p^2\over 2m}+{e^2\over r}\sim \bar H+{\l^2\over 2}\D H,\eqno(18)$$
with
$$\D H={\br^2\bp^2\over m}-e^2\br.\eqno(19)$$
The energy spectrum of the unperturbed Hamiltonian is given by
$$E_n=-{me^4\over 2n^2}\eqno(20)$$
In analogy with the harmonic oscillator, one can write
$${\bp^2\over 2m}=\bar H-{e^2\over\br},\eqno(21)$$
obtaining a diagonal matrix with (again the quantum number $m$ is not relevant)
$$<n,l\,|\D H|\,n,l>\ =2E_n<n,l\,|\,\br^2|\,n,l>-\ 3e^2<n,l\,|\,\br|\,n,l>.\eqno(22)$$
Using [10]
$$<n,l\,|\,\br|\,n,l>\ ={1\over2me^2}\,[3n^2-l(l+1)],$$
$$<n,l\,|\,\br^2|\,n,l>\ ={n^2\over2m^2e^4}\,[5n^2-3l(l+1)+1],$$
one gets
$$\D E_{n,l}=-{1\over m}\left[7n^2-3l(l+1)+\ha\right].\eqno(23)$$
Hence, for $l=0$, the relative strength of the corrections is
$${\l^2\over2}\,{\D E_{n,0}\over E_n}\sim{7n^4\l^2\over2m^2e^4}={7\over2}n^4a^2\l^2,\eqno(24)$$
where $a$ is the Bohr radius. The corrections are therefore of the order of the square
of the ratio between the Bohr radius and the de Sitter radius $1/\l$,
i.e.\ $10^{-72}$. This effect is even tinier than the one due to the generalized
uncertainty principle, and could be detectable experimentally only if the parameter
$\l^2$ in our formulae were much bigger than the observed value of the cosmological
constant.
\bigbreak

\section{3. Lamb shift}

A slightly different calculation can be performed to obtain corrections
to the Lamb shift effect in the hydrogen atom, in analogy with the investigation
of ref.\ [5] in the case of the generalized \up.
The shift in the wave function is at first order [10]
$$\q_{nlm}=\bar\q_{nlm}+\sum_{n\ne n'}{<n',l,'m'|\D H|\,n,l,m>\over E_n-E_{n'}}\
\bar\q_{n'l'm'},\eqno(25)$$
where a bar indicates the unperturbed wave function.
Using the standard expression of the wave function and the orthogonality relations
of the spherical harmonics, the shift in the ground state wave function yields
$$\D\q_{100}=\q_{100}-\bar\q_{100}={<2,0,0\,|\D H|\,1,0,0>\over E_1-E_2}\ \bar\q_{200}.
\eqno(26)$$
Substituting the expression (19) for $\D H$, the explicit calculation of the matrix
element gives
$$<2,0,0|\D H|1,0,0>\ =\ <2,0,0|\,H\br^2+\br^2H-3e^2\br\,|1,0,0>\ =
{608\sqrt2\over243}\,ae^2,\eqno(27)$$
where we have used a symmetric ordering for the first term.
It follows that
$$\D\q_{100}={2432\sqrt2\over729}\,\l^2a^2\bar\q_{200}.\eqno(28)$$
On the other hand, the Lamb shift for the ground state of the hydrogen atom
is given by
$$\D E_1=-{4e^2\ln(e^2)\over3m^2}|\q_{100}(0)|^2.\eqno(29)$$
The contribution due to the modification of the commutation relations is therefore
$${\D E_1^{EUP}\over\D E_1}={2|\D\q_{100}(0)|\over\q_{100}(0)}={2432\over 729}\,\l^2a^2.
\eqno(30)$$
Also in this case the effect is of order $\l^2a^2$ and hence not detectable experimentally.

\section{4. Conclusions}

The corrections to the spectra of quantum mechanical systems due to the EUP are qualitatively
similar to those associated to the GUP, discussed in ref.\ [3-5].
However, their size is different. While those
associated to GUP are of relative size $(l_P/L)^2$, $L$ being the typical length scale
of the system, in the case of EUP they are of order $(\l\,L)^2$, as could have been expected
from dimensional arguments. For systems like the hydrogen atom they are of order $10^{-48}$
and $10^{-72}$ respectively, and therefore much smaller in the EUP case. Of course they are
not detectable experimentally, unless some mechanism which fixes a value for $\l$ in the
\up greater than the one associated with the cosmological constant is available.
The same calculations hold in the case of anti-de Sitter spacetime, but the corrections
have opposite sign.

The main result of this paper is that it is possible, at least in principle, to detect effects
due to the large scale curvature of spacetime in atomic experiments. However, with the accuracy
of present day experimental setup, this  is still of theoretical interest.
\vspace{.5cm}
\begin{center}
{\bf References}
\end{center}
\noindent\ref [1] D. Amati, M. Ciafaloni and G. Veneziano, \PL{B216}, 41 (1989);
M. Maggiore, \PL{B304}, 63 (1993).\\
\ref [2] M.I. Park, \PL{B659}, 698 (2008).\\
\ref [3] A. Kempf, \JoP{A30}, 2093 (1999).\\
\ref [4] F. Brau, \JoP{A32}, 7691 (1999).\\
\ref [5] S. Das and E.C. Vagenas, \PRL{101}, 221301 (2008).\\
\ref [6] B. Bolen and M. Cavagli\`a, \GRG{37}, 1255 (2005).\\
\ref [7] C. Bambi and F.R. Urban, \CQG{25}, 095006 (2008).\\
\ref [8] S. Mignemi, \MPL{A25}, 1697 (2010).\\
\ref [9] Similar transformations have been employed in other forms of noncommutative
spacetime, see for example S. Ghosh and P. Pal, \PR{D75}, 105021 (2007).\\
\ref [10] A. Messiah, {\it Quantum Mechanics}, Dover 1999.

\end{document}